\documentclass[aps,prc,twocolumn,superscriptaddress,showpacs]{revtex4-1}

\usepackage{graphicx}% Include figure files
\usepackage{dcolumn}% Align table columns on decimal point
\usepackage{bm}% bold math
\usepackage{ulem}
\usepackage{color}

\newcommand{\al}{$\alpha$}

\newcommand{\raa}{($\alpha$,$\alpha$)}

\newcommand{\rag}{($\alpha$,$\gamma$)}

\newcommand{\rap}{($\alpha$,$p$)}
\newcommand{\rpa}{($p$,$\alpha$)}

\newcommand{\rpg}{(p,$\gamma$)}

\newcommand{\rpt}{($p$,$t$)}
\newcommand{\rpp}{($p$,$p$)}

\newcommand{\Nsv}{$N_A$$\left< \sigma v \right>$}
\newcommand{\sfact}{S-factor}

\begin{document}

\title{
  Thermonuclear reaction rate of 
  $^{18}$Ne($\alpha$,$p$)$^{21}$Na from Monte-Carlo calculations
}

\author{P.\ Mohr}
\email[Email: ]{WidmaierMohr@t-online.de}
\affiliation{
Diakonie-Klinikum, D-74523 Schw\"abisch Hall, Germany}
\affiliation{
Institute for Nuclear Research (ATOMKI), H-4001 Debrecen, Hungary}

\author{R.\ Longland}
%\altaffiliation[Present address: ]{}
%\affiliation{Department of Physics and Astronomy, University of North
%  Carolina, Chapel Hill, NC 27599-3255, USA}
\affiliation{Department of Physics, North Carolina State University, Raleigh,
  NC 27695-8202, USA} 
\affiliation{Triangle Universities Nuclear Laboratory, Durham, NC 27708-0308,
  USA}

\author{C.\ Iliadis}
%\altaffiliation[Present address: ]{}
\affiliation{Department of Physics and Astronomy, University of North
  Carolina, Chapel Hill, NC 27599-3255, USA}
\affiliation{Triangle Universities Nuclear Laboratory, Durham, NC 27708-0308,
  USA}

\date{\today}

\begin{abstract}
The $^{18}$Ne($\alpha$,$p$)$^{21}$Na reaction impacts the break-out from
the hot CNO-cycles to the $rp$-process in type I X-ray bursts. We present a revised thermonuclear reaction
rate, which is based on the latest experimental
data. The new rate is derived from Monte-Carlo calculations, taking into
account the uncertainties of all nuclear physics input quantities. In addition, we
present the reaction rate uncertainty and probability density versus temperature.
Our results are also consistent with estimates obtained using different indirect approaches.
\end{abstract}

% 25.60.-t 	Reactions induced by unstable nuclei
% 25.55.-e 	3H-, 3He-, and 4He-induced reactions
% 26.30.-k 	Nucleosynthesis in novae, supernovae and other explosive environments

\maketitle

\section{Introduction}
\label{sec:intro}
There has been much interest recently in the thermonuclear rate of the
$^{18}$Ne\rap$^{21}$Na reaction at temperatures of type I X-ray bursts
\cite{Sch06,Par08}. Because of the high temperatures of several
Giga-Kelvin (in usual notation: $T_9 \approx 1-2$), the reaction rate is essentially defined by the cross section at
energies between 1 and 3\,MeV. This corresponds to excitation energies of
$E^\ast \approx 9-11$\,MeV in the compound nucleus $^{22}$Mg.

The latest studies have focused on indirect determinations
of the $^{18}$Ne\rap $^{21}$Na reaction rate. Matic {\it et al.}\ \cite{Matic09} 
obtained excitation energies, $E^\ast$, of many levels in the $^{22}$Mg compound nucleus
 by measuring the $^{24}$Mg\rpt $^{22}$Mg reaction. These energies define the
resonance energies, $E$, in the $^{18}$Ne\rap $^{21}$Na reaction, which enter
exponentially into the expression for the reaction rate and are thus
the most important ingredient.
From the same experiment, total widths $\Gamma$ of these states were
derived \cite{Mohr13}. In addition, spins and parities, $J^\pi$, of several
states in $^{22}$Mg have been measured recently by resonant proton scattering
using the $^{21}$Na\rpp $^{21}$Na reaction in inverse kinematics
\cite{He13,Zhang14}. Furthermore, the reverse $^{21}$Na\rpa $^{18}$Ne
reaction has been
used in two independent experiments \cite{Sal12,ANL} to determine a lower limit of the forward $^{18}$Ne\rap $^{21}$Na
reaction cross section. Mohr and Matic~\cite{Mohr13} found a dramatic disagreement between
the earlier forward $^{18}$Ne\rap $^{21}$Na reaction data obtained by Groombridge
{\it et al.}\ \cite{Gro02} and the reverse reaction data \cite{Sal12,ANL}.
Consequently, the data of Ref.~\cite{Gro02} were excluded from the
determination of the reaction rate in Ref.~\cite{Mohr13}.

Two different approaches have been used in Ref.~\cite{Mohr13} to determine the $^{18}$Ne\rap $^{21}$Na reaction rate from the available data. In the first
approach, the experimental resonance energies from the $^{24}$Mg\rpt $^{22}$Mg
reaction \cite{Matic09}, together with \al -transfer data for the mirror compound
nucleus $^{22}$Ne \cite{Matic09}, were employed for calculating the required resonance strengths, $\omega \gamma_{\alpha p}$.
In the second approach, the experimental reverse reaction
data \cite{Sal12,ANL} were corrected for thermal target excitations by using a Hauser-Feshbach model and the forward rate was obtained using the reciprocity theorem.
It was shown in
Ref.~\cite{Mohr13} that the reaction rates obtained by these two methods differ by a factor of $\approx$3. Taking
into account estimated uncertainties of a factor of $\approx$2 for both
approaches, the geometric mean value has been recommended in Ref.~\cite{Mohr13}.

The present study involves several major improvements compared to the procedure of Ref.~\cite{Mohr13}. We employ
the latest $J^\pi$ assignments from resonant $^{21}$Na\rpp $^{21}$Na
elastic scattering \cite{He13,Zhang14}. Furthermore, for all states seen in
the $^{24}$Mg\rpt $^{22}$Mg transfer experiment \cite{Matic09}, 
improved resonance strengths, $\omega \gamma_{\alpha p}$, are determined (Sec.~\ref{sec:strength}). The new
resonance strengths are used as input for a Monte-Carlo sampling method 
\cite{Long10,Ili10a,Ili10b,Ili10c}, which provides for the first time the rate probability densities 
of the $^{18}$Ne\rap $^{21}$Na reaction. From these results we extract statistically meaningful
thermonuclear rates and rate uncertainties (Sec.~\ref{sec:MC}). Finally,
the break-out temperature and its uncertainty are given for typical
conditions of type I X-ray bursts.

\section{Preliminary considerations}
\label{sec:basic}
The reaction rate,
\Nsv , of the $^{18}$Ne\rap $^{21}$Na reaction is given by the sum over the
contributions of many resonances. (Strictly speaking, the reaction rate is
obtained after multiplying \Nsv\ with the densities of the reaction partners
and integration over space. Nevertheless, the quantity \Nsv\ is usually called
``reaction rate''. We keep this convention throughout this paper.)

\begin{figure}[htb]
\includegraphics[width=\columnwidth,clip=]{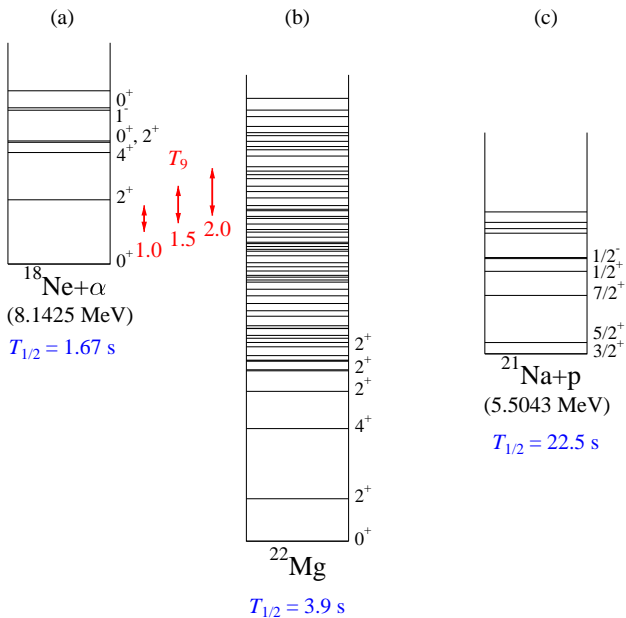}
\caption{
\label{fig:level}
(Color online)
Level scheme of the compound nucleus $^{22}$Mg (b) with the $^{18}$Ne+\al\ (a)
and $^{21}$Na+$p$ (c) thresholds (approximately to scale). The energies are
given in MeV. The energy range of the classical Gamow window is indicated for
three temperatures, $T_9 = 1.0$, 1.5, and 2.0 by red arrows. 
}
\end{figure}

A simplified level scheme for the $^{18}$Ne\rap $^{21}$Na reaction
is shown in Fig.~\ref{fig:level} for illustration
purposes. Because of the positive $Q$-value of this 
reaction, the available energy in the $^{21}$Na+$p$ channel is much higher
than in the $^{18}$Ne+\al\ channel. Simultaneously, the Coulomb barrier is
much lower in the $^{21}$Na+$p$ channel ($Z_1 \times Z_2 = 11$) compared to
the $^{18}$Ne+\al\ channel ($Z_1 \times Z_2 = 20$). Hence, the proton partial
width is much larger compared to the $\alpha$-particle partial width for all
relevant resonances, $\Gamma_p 
\gg \Gamma_\alpha$, and the $\gamma$-ray partial width, $\Gamma_\gamma$, is
very small 
compared to the total width, $\Gamma_p \approx \Gamma$. 

It has been shown in Ref.~\cite{Mohr13} that the reaction rate can be
calculated using the narrow resonance formalism \cite{Ili07}:
\begin{equation}
N_A \left< \sigma v \right>
 = 
\frac{1.54 \times 10^{11}}{(\mu \, T_9)^{3/2}}  
\sum_i \frac{(\omega \gamma_{\alpha p})_i}{e^{11.605 E_i/T_9}}~{\rm{cm}}^3 \, {\rm{s}}^{-1} \, {\rm{mol}}^{-1}
\label{eq:rate}
\end{equation}
with the reduced mass, $\mu$, in units of amu, the resonance energies,
$E_i$, in MeV, and the resonance strengths, $(\omega \gamma_{\alpha p})_i$, in MeV. In
this work, resonance energies and excitation energies are denoted by $E$ and
$E^\ast$, respectively. All quantities are given in the center-of-mass (c.m.)
system unless noted otherwise. The deviations between the simple
narrow-resonance formalism and a numerical integration of the cross section
$\sigma(E)$ in \Nsv\ remain below 5\,\% for $T_9 = 1-2$ and below 10\,\% for a
wider temperature range of $T_9 = 0.25 - 3$ \cite{Mohr13}.

Resonance energies are derived from the measured excitation energies
\cite{Matic09} by using $E = E^\ast - S_\alpha$, with the \al -particle
binding energy in the 
$^{22}$Mg compound nucleus given by
$S_\alpha = 8142.5 \pm 0.5$~keV \cite{AMDC}. The uncertainty of $S_\alpha$ does not significantly affect the uncertainty of the reaction rate.
The resonance energies are thus well constrained by the
high-resolution transfer experiment of Ref.~\cite{Matic09}. Earlier
transfer experiments \cite{Chen01,Cagg02,Berg03,Chae09}
show good agreement with the high-resolution data \cite{Matic09}
(for a detailed discussion, see Ref.~\cite{Mohr13}).

The resonance strength, $\omega \gamma_{\alpha p}$, for a resonance with spin $J$ in the
$^{18}$Ne\rap $^{21}$Na reaction is given by
\begin{equation}
\omega \gamma_{\alpha p} = (2J+1) \, \frac{\Gamma_\alpha \Gamma_p}{\Gamma} \approx (2J+1)
\Gamma_\alpha 
\label{eq:strength}
\end{equation}
where the approximations $\Gamma_p \approx \Gamma$ and $\Gamma_\alpha
\ll \Gamma_p$ are used in Eq.~(\ref{eq:strength}). Thus, the essential
quantity defining the resonance strength is the partial width
$\Gamma_\alpha$ for the decay of an excited level into the $^{18}$Ne+\al\ channel.
The estimation of $\Gamma_\alpha$ for $^{22}$Mg levels from properties of
the mirror states in $^{22}$Ne will be described in detail in
Sec.~\ref{sec:strength}. 

The first excited state in $^{18}$Ne with $J^\pi = 2^+$ is located at a
relatively high excitation energy of $E^\ast = 1887$\,keV. In a stellar plasma
the population of the first excited state remains negligible with
$(2J+1) \exp{(-E^\ast/kT)} \lesssim 10^{-4}$ up to temperatures $T_9 < 2$, and
the contributions of higher-lying states are even smaller. Thus, 
at typical temperatures of type I X-ray bursts
the {\it stellar} rate of the $^{18}$Ne\rap $^{21}$Na reaction 
(i.e., including thermal target excitations)
is practically identical to the {\it laboratory} rate
(i.e., for target ground state population only).

The situation is different for the reverse $^{21}$Na\rpa $^{18}$Ne
reaction. Low-lying states in $^{21}$Na may be populated under stellar
conditions, and a laboratory measurement of the $^{21}$Na\rpa $^{18}$Ne
reaction cannot determine the stellar $^{21}$Na\rpa $^{18}$Ne rate. The
laboratory measurement provides only the partial $^{21}$Na$_{\rm{g.s.}}$\rpa
$^{18}$Ne$_{\rm{g.s.}}$ cross section. This statement holds exactly for low
proton energies below 4.5\,MeV where the reaction channel to the first excited
state in $^{18}$Ne is closed. But also at slightly higher energies this
statement holds approximately because the Coulomb
barrier strongly prefers the ground state channel, and no event for the first
excited state has been detected in the experiment of Salter {\it et al.}
\cite{Sal12}. 

From the above considerations we may conclude that a measurement of the
reverse $^{21}$Na\rpa $^{18}$Ne reaction only provides the
$^{18}$Ne$_{\rm{g.s.}}$\rap $^{21}$Na$_{\rm{g.s.}}$ cross section using the
reciprocity theorem of nuclear reactions. A reaction rate which is determined
from this ground state to ground state cross section is only a lower limit for
the stellar rate of the forward $^{18}$Ne\rap $^{21}$Na reaction which
populates the ground state and low-lying states of $^{21}$Na.

For completeness we mention that {\it stellar} reaction rates of forward and
reverse reactions are directly related by detailed balance. However, this
general relation does not hold for rates which are determined from laboratory
cross sections.

The thermonuclear rate of the
$^{18}$Ne\rap $^{21}$Na reaction covers many orders of magnitude in the relavant
temperature range. Therefore, it is helpful to compare the present results to a reference rate. For the latter we adopt the rate of Ref.~\cite{Mohr13}, which is based on experimental resonance energies and total widths and on calculated
$\alpha$-particle partial widths and resonance strengths (see Tables I and II
of Ref.~\cite{Mohr13}). 

\section{Determination of resonance strengths}
\label{sec:strength}
The partial widths $\Gamma_\alpha$ of $^{22}$Mg levels in the relevant energy
window have not been measured directly yet. Hence indirect methods have to be applied to
determine $\Gamma_\alpha$. In such cases, the assumption of mirror symmetry in the wave functions of corresponding $^{22}$Mg and $^{22}$Ne 
levels yields for the dimensionless reduced widths
\begin{equation}
\theta_\alpha^2(^{22}{\rm{Mg}}) \approx \theta_\alpha^2(^{22}{\rm{Ne}})
\label{eq:reduced}
\end{equation}
Equation~(\ref{eq:reduced}) is typically fulfilled for states
with a strong \al -cluster component in the wave function (i.e., for a large value of
$\theta_\alpha^2$), whereas significant discrepancies have been found for
mirror states with very small $\theta_\alpha^2$ values \cite{Oli97}.

Here we follow the procedure for estimating $\Gamma_\alpha$ in $^{22}$Mg that was outlined in Ref.~\cite{Matic09}, but implement several improvements. We group the astrophysically important $^{22}$Mg levels, listed in Refs.~\cite{Matic09,Mohr13} into four different categories, which will be discussed below in more detail.

\subsection{Resonance strengths from reduced widths in the $^{22}$Ne mirror
  nucleus}
\label{sec:mirror}
{\it Category A} represents states with known $\Gamma_\alpha$ in the mirror nucleus $^{22}$Ne. Absolute $\Gamma_\alpha$ values can be determined
from $^{18}$O\rag $^{22}$Ne reaction data for a few low-lying resonances
\cite{Tra78,Vog90,Gie94,Dab03} and from resonant $^{18}$O\raa $^{18}$O
elastic scattering at higher energies \cite{Gol04}. However, the latter do not
play a significant role in the determination of \Nsv\ 
because of the relatively high energies. For states in this category A, we
assign an 
uncertainty of a factor of two to the derived partial widths, $\Gamma_\alpha$,
in $^{22}$Mg. The corresponding $\Gamma_\alpha$ values in $^{22}$Ne are typically known
with much smaller uncertainties. The factor of two uncertainty reflects
the assumption of mirror symmetry of the wave functions in
Eq.~(\ref{eq:reduced}).

{\it Category B} consists of a few high-lying states, corresponding to resonance energies above 3.5 MeV, with known
$\Gamma_\alpha$ values in the $^{22}$Ne mirror nucleus from resonant $^{18}$O\raa
$^{18}$O elastic scattering \cite{Gol04}. Because of their high resonance energy, the partial widths $\Gamma_\alpha$ in
$^{22}$Mg become very large, and the approximation
$\Gamma_\alpha \ll \Gamma$ does not hold anymore. In this case the
resonance strengths, $\omega \gamma_{\alpha p} = \omega \Gamma_\alpha \Gamma_p / \Gamma$,
are determined using $\Gamma_p = \Gamma - \Gamma_\alpha$ where the total
widths $\Gamma$ are taken from \cite{Mohr13} and the partial widths
$\Gamma_\alpha$ are calculated from Eq.~(\ref{eq:reduced}) and
$\theta_\alpha^2$ from \cite{Gol04}. 
For states in this
category we assign a resonance strength uncertainty of a factor 3. Note that
these levels have practically no impact on \Nsv\ at astrophysically relevant
temperatures.

{\it Category C} is assigned to states with known spectroscopic information from
\al -transfer on $^{18}$O. The relevant excitation energy range was
studied in Ref.~\cite{Gie94} using the $^{18}$O($^6$Li,$d$)$^{22}$Ne reaction at
$E_{\rm{lab}} = 32$\,MeV. Because only relative spectroscopic factors are
reported in Ref.~\cite{Gie94}, an absolute normalization of the transfer data has
to be performed (see below). For states in this category, we 
assign a resonance strength uncertainty of a factor 3, which is a combined
uncertainty of the assumption of mirror symmetry in Eq.~(\ref{eq:reduced}) and
the model dependence introduced by deriving $\Gamma_\alpha$ from \al
-transfer. The estimate of a factor of two uncertainty for the assumption of
mirror symmetry was already explained for the Category A states above. We
estimate another factor of two uncertainty for the model dependence of the
determination of reduced widths $\theta_\alpha^2$ and absolute partial widths
$\Gamma_\alpha$ from transfer. Combining both
uncertainty factors using quadratic error propagation for lognormal
distributions, leads to an overall uncertainty factor of $\sqrt{2^2 + 2^2} =
\sqrt{8} \approx 3$.

{\it Category D} is assigned to all of the remaining levels for which no spectroscopic information is available.
As a crude estimate, Matic {\it et al.}~\cite{Matic09} adopted for these states a reduced width that was obtained by averaging the experimental values of levels observed in transfer studies. This procedure is problematic because transfer reactions preferentially populate states with large
reduced widths, $\theta_\alpha^2$. In other words, states that have not been observed in transfer reactions
presumably have smaller reduced widths compared to the average of the detected
states. Therefore, the procedure applied by Ref.~\cite{Matic09} significantly overestimates the actual dimensionless reduced width, $\theta_\alpha^2$, and thus the $\alpha$-particle partial width, $\Gamma_\alpha$. 
In this work, we randomly sample $\theta_\alpha^2$ for all states in Category D from a
Porter-Thomas distribution with a mean reduced width of $\left< \theta_\alpha^2 \right> = 0.03
\pm 0.01$. We obtain this value by extrapolating the recent results of Pogrebnyak {\it et
  al.}~\cite{Pog13} for nuclei with slightly larger mass numbers $A$. In the latter work, 
$\left< \theta_\alpha^2 \right> = 0.018$ was found with a trend to increasing values for
smaller mass numbers. The value of $\left< \theta_\alpha^2 \right> = 0.03
\pm 0.01$ used in the present work leads to a
significantly reduced reaction rate compared to the results of Ref.~\cite{Matic09}, as shown below.
We also find that the uncertainty in this value has only a minor impact on the total rate. 
A further uncertainty for \Nsv\ arises from only tentative $J^\pi$
  assignments for these category D states without detailed spectroscopic
  information. This will be discussed later (see Sect.~\ref{sec:MC}).

The partial width $\Gamma_\alpha$ is calculated from the single-particle
($s.p.$) limit according to
\begin{equation}
\Gamma_\alpha = \theta_\alpha^2 \times \Gamma_\alpha^{s.p.}
\label{eq:wigner}
\end{equation}
where $\Gamma_\alpha^{s.p.}$ is computed using a radius of
$R = R_0 \times (A_P^{1/3} + A_T^{1/3})$, with $R_0 = 1.25$\,fm. This value of the radius parameter is chosen for consistency with Ref.~\cite{Pog13}. The choice
of $R_0$ has only a small influence on the calculated $\Gamma_\alpha$ values in
$^{22}$Mg because the same $R_0$ has been used in the determination of
$\theta_\alpha^2(^{22}{\rm{Ne}})$ from $\Gamma_\alpha(^{22}{\rm{Ne}})$ and in
the determination of $\Gamma_\alpha(^{22}{\rm{Mg}})$ from
$\theta_\alpha^2(^{22}{\rm{Mg}}) \approx \theta_\alpha^2(^{22}{\rm{Ne}})$.

Apart from the above modification in the treatment of states without
spectroscopic information 
(category D), we implement two other improvements compared to earlier work 
\cite{Matic09,Mohr13}. 

First, for some states very small reduced widths, on the order of $10^{-5}$, were
reported in Ref.~\cite{Matic09}, which were derived from 
$^{18}$O\rag $^{22}$Ne capture data \cite{Tra78}. However, for these
resonances the neutron channel is also open, and it is not possible to
derive $\Gamma_\alpha$ from the measured \rag\ resonance strength, $\omega
\gamma_{\alpha\gamma} = (2J+1) \Gamma_\alpha \Gamma_\gamma / \Gamma \ne (2J+1) \Gamma_\alpha$,
because the total width, $\Gamma$, may be dominated by the neutron width,
$\Gamma_n$. We have assigned these states either to category C or category D,
depending on whether reduced widths were measured in the transfer experiment \cite{Gie94}.

Second, as already pointed out above, only relative spectroscopic factors are available for
the states assigned to category C. Matic {\it et al.}~\cite{Matic09} normalized the
transfer data to theoretical spectroscopic factors, either of the ground state
or of the state at $E^\ast = 10066$\,keV. In the present work we instead normalize 
the spectroscopic factors to experimental values only. The $1^-$ state in the
$^{22}$Ne mirror nucleus at $E^\ast = 10209$\,keV has been detected in the
$^{18}$O($^6$Li,$d$)$^{22}$Ne transfer experiment \cite{Gie94} and in
several $^{18}$O\rag $^{22}$Ne capture experiments
\cite{Tra78,Vog90,Gie94}. The capture results 
are in excellent agreement. The most recent capture experiment \cite{Dab03}
recommends an experimental resonance strength of $\omega \gamma_{\alpha\gamma} = 229 \pm
19$\,$\mu$eV, which was also used to determine the strengths of lower-lying observed resonances. From this experimental value, an $\alpha$-particle partial width and a reduced width of 
 $\Gamma_\alpha = \omega \gamma/3 =76.3$\,$\mu$eV and 
$\theta_\alpha^2 = 0.150 \pm 0.012$, respectively, can be derived using $R_0 =
1.25$\,fm. Giesen {\it et al.}~\cite{Gie94} report for this level, which is located below the neutron threshold in $^{22}$Ne, a relative
spectroscopic factor of $S_{\rm{rel}} = 0.035$ from an
$^{18}$O($^6$Li,$d$)$^{22}$Ne experiment. Consequently, we scaled all relative spectroscopic factors of Ref.~\cite{Gie94} by a factor $f $ = $0.150 / 0.035$ = $4.29$ in order to obtain
the reduced width, $\theta_\alpha^2$. Although the uncertainty in $f$ is small, we assigned 
a factor of 3 uncertainty to all resonance strengths of category C levels that involved an estimate of $\Gamma_\alpha$ in $^{22}$Mg using this scaling factor.

Our recommended resonance strengths, $\omega \gamma_{\alpha p}$, are listed in
Table \ref{tab:str}. For states in category D we provide $\omega \gamma$ in
parenthesis which are calculated from $\Gamma_\alpha = 0.03 \times
\Gamma_\alpha^{s.p.}$ where the factor of 0.03 is taken from the systematics
of reduced widths \cite{Pog13}. Note that these resonance strengths are used in
the presentation of the recommended astrophysical \sfact\ in the next paragraph;
however, these numbers do not enter directly in the calculation of
\Nsv\ because here a Porter-Thomas distribution will be sampled using the
Monte-Carlo method. Additionally, the previous values, $\omega
\gamma_{\rm{ref}}$, from Tables I and II of Ref.~\cite{Mohr13} are provided
for comparison.
\begin{table*}[tbh]
\caption{\label{tab:str}
Properties of resonances in $^{18}$Ne\rap $^{21}$Na; the new
resonance strengths $\omega \gamma_{\alpha p}$ from this work are compared to the
reference strengths $\omega \gamma_{\rm{ref}}$ from
Ref.~\cite{Mohr13}. Excitation energies in the compound nucleus 
$^{22}$Mg and the mirror assignments in $^{22}$Ne are given. All reference
values are adopted from Ref.~\cite{Mohr13}. Dimensionless reduced widths
labeled ``PT'' indicate states of category D; for the calculation of the
reaction rate $\Gamma_\alpha$ is sampled according to a Porter-Thomas
distribution with $\left< \theta_\alpha^2 \right>$ = $0.03 \pm 0.01$
(see the text for details). For these states we provide
$\omega \gamma_{\alpha p} \approx (2J+1) \Gamma_\alpha = (2J+1) \times
0.03 \times \Gamma_\alpha^{s.p.}$ in
parenthesis. 
}
%\begin{center}
\begin{tabular}{ccrr@{$\pm$}lr@{$\times$}llr@{$\times$}lccc}
\hline
\multicolumn{1}{c}{$E^\ast(^{22}{\rm{Mg)}}$ (MeV)}
& \multicolumn{1}{c}{$E$ (MeV)}
& \multicolumn{1}{c}{$J^\pi_{\rm{ref}}$}
& \multicolumn{2}{c}{$\Gamma_{\rm{ref}}$ (keV)}
& \multicolumn{2}{c}{$\omega \gamma_{\rm{ref}}$ (eV)}
& \multicolumn{1}{c}{$J^\pi$}
& \multicolumn{2}{c}{$\omega \gamma$ (eV)}
& \multicolumn{1}{c}{$\theta_\alpha^2$}
& \multicolumn{1}{c}{category}
& \multicolumn{1}{c}{$E^\ast(^{22}{\rm{Ne)}}$ (MeV)} \\
\hline
  8.182 & 0.040 & $[2^+]$ &  33.5  &    2.2 & 8.53 & $10^{-65}$ 
  & $2^+$ \footnote{$J^\pi = 2^+$ confirmed in Ref.~\cite{Zhang14}}  
           & (3.30 & $10^{-66}$) & PT      & D & 8.489 \\
  8.385 & 0.243 & $[2^+]$ &  47.0  &    5.3 & 1.33 & $10^{-17}$ 
  & $1^+$ \footnote{new $J^\pi = 1^+$ assignment in Ref.~\cite{Zhang14}} 
           & \multicolumn{2}{c}{--}  & --      & -- & \\
  8.519 & 0.377 & $[2^+]$ &  25.7  &    4.1 & 1.53 & $10^{-11}$
  & $3^-$ \footnote{earlier $J^\pi = 3^-$ of \cite{Matic09} confirmed; $J^\pi
    = 2^+$ of Ref.~\cite{Chae09} rejected in Ref.~\cite{Zhang14}} 
           & 7.05 & $10^{-14}$   & 0.017   & C & 8.740 \\
  8.574 & 0.432 & $[4^+]$ &  20.6  &   16.8 & 3.26 & $10^{-12}$
  & $4^+$ \footnote{$J^\pi = 4^+$ confirmed in Ref.~\cite{Zhang14}} 
           & (5.05 & $10^{-13}$) & PT      & D & 8.855 \\
  8.657 & 0.515 & $[0^+]$ &  15.5  &    3.5 & 4.97 & $10^{-8}$
  & $2^+$ \footnote{new $J^\pi = 2^+$ assignment in Ref.~\cite{Zhang14}} 
           & (3.13 & $10^{-9}$)  & PT      & D & 8.596 \\
  8.743 & 0.601 & $[4^+]$ &  65.5  &   22.8 & 5.15 & $10^{-9}$ 
  & $2^+$ \footnote{new $J^\pi = 2^+$ assignment in Ref.~\cite{Zhang14}} 
           & (1.23 & $10^{-7}$)  & PT      & D & 9.045 \\
  8.783 & 0.641 & $[1^-]$ &  22.5  &    7.0 & 1.21 & $10^{-5}$ 
  & $1^-$ \footnote{$J^\pi = 1^-$ confirmed in Ref.~\cite{Zhang14}}  
           & (1.32 & $10^{-6}$)  & PT      & D & 9.097 \\
  8.932 & 0.790 & $[2^+]$ &  51.6  &    5.9 & 4.13 & $10^{-4}$ 
  & $2^+$ \footnote{$J^\pi = 2^+$ confirmed in Ref.~\cite{Zhang14}}  
           & (4.14 & $10^{-5}$)  & PT      & D & 9.229 \\
  9.080 & 0.938 & $[1^-]$ & 114.4  &   19.7 & 2.31 & $10^{-2}$ 
  & $1^-$ \footnote{$J^\pi = 1^-$ confirmed in Ref.~\cite{Zhang14}}  
           & 5.25 & $10^{-3}$    & 0.064   & C & 9.324 \\
  9.157 & 1.015 & $[4^+]$ & \multicolumn{2}{c}{$< 20.5$} & 8.70 & $10^{-4}$ 
  & $4^+$ \footnote{$J^\pi = 4^+$ confirmed in Ref.~\cite{Zhang14}}  
           & (1.01 & $10^{-4}$)  & PT      & D & 9.508 \\
  9.318 & 1.176 & $[2^+]$ &  22.6  &    8.0 & 4.97 & $10^{-1 }$ 
  & $2^+$  & (4.97 & $10^{-2}$)  & PT      & D & 9.625 \\
  9.482 & 1.340 & $[3^-]$ & \multicolumn{2}{c}{$< 6.3$} & 1.25 & $10^{-1}$ 
  & $3^-$  & 1.35 & $10^{-1}$    & 0.047   & C & 9.725 \\
  9.542 & 1.400 & $[1^-]$ & \multicolumn{2}{c}{$< 22.9$} & 1.31 & $10^{1}$
  & $1^-$  & 5.74 & $10^{0}$     & 0.115   & C & 9.842 \\
  9.709 & 1.567 & $[0^+]$ & 267.8  &   48.2 & 5.18 & $10^{1}$ 
  & $0^+$  & 6.50 & $10^{1}$     & 0.458   & A & 10.052 \\
  9.752 & 1.610 & $[1^-]$ &  31.4  &    6.8 & 4.82 & $10^{1}$ 
  & $2^+$ \footnote{$J^\pi = 2^+$ from $J^\pi = (1^-,2^+)$ in Ref.~\cite{Chae09}
    and $J^\pi = (2^+, 3^-, 4^+)$ in Ref.~\cite{Gie94}}   
           & 8.45 & $10^{0}$     & 0.052   & A & 10.137 \\
  9.860 & 1.718 & $[0^+]$ & 121.3  &   10.4 & 2.07 & $10^{1}$
  & $0^+$  & 1.92 & $10^{1}$     & 0.042   & A & 10.283 \\
 10.085 & 1.943 & $[2^+]$ &  25.8  &    9.3 & 2.25 & $10^{2}$ 
  & $2^+$  & 2.34 & $10^{2}$     & 0.134   & A & 10.297 \\
 10.272 & 2.130 & $2^+$   &  20.7  &    2.7 & 1.31 & $10^{3}$ 
  & $2^+$  & (1.49 & $10^{2}$)   & PT      & D & 10.551 \\
 10.429 & 2.287 & $[4^+]$ & 144.2  &   25.8 & 4.89 & $10^{1}$ 
  & $1^-$ \footnote{$J^\pi = 1^-$ from Ref.~\cite{Chae09}}
           & 9.39 & $10^{2}$     & 0.150   & A & 10.209 \\
 10.651 & 2.509 & $[3^-]$ &  72.8  &   19.1 & 1.12 & $10^{3}$
  & $3^-$  & (2.65 & $10^{2}$)   & PT      & D & 10.857 \\
 10.768 & 2.626 & $[2^+]$ &  94.9  &   29.6 & 1.16 & $10^{4}$ 
  & $2^+$  & (1.29 & $10^{3}$)   & PT      & D & 11.064 \\
 10.873 & 2.731 & $[0^+]$ &  40.2  &   12.0 & 1.19 & $10^{4}$
  & $0^+$  & (1.59 & $10^{3}$)   & PT      & D & 11.194 \\
 11.001 & 2.859 & $[4^+]$ & 135.8  &   12.9 & 5.81 & $10^{2}$ 
  & $4^+$  & (2.13 & $10^{2}$)   & PT      & D & 11.271 \\
 11.315 & 3.173 & $[4^+]$ & 203.7  &   37.0 & 1.83 & $10^{3}$
  & $4^+$  & (6.21 & $10^{2}$)   & PT      & D & 11.577 \\
 11.499 & 3.357 & $[2^+]$ & 116.8  &   21.8 & 8.64 & $10^{4}$ 
  & $2^+$  & 2.17  & $10^{4}$    & 0.062   & A & 11.700 \footnote{$E^\ast$,
   $J^\pi$, and $\Gamma_\alpha$ adopted from Ref.~\cite{Gol04}} \\
 11.595 & 3.453 & $[4^+]$ &  48.3  &   14.7 & 3.67 & $10^{3}$
  & $4^+$  & (1.41 & $10^{3}$)   & PT      & D & 11.896 \\
 11.747 & 3.605 & $[0^+]$ & 166.1  &   64.4 & 7.13 & $10^{4}$ 
  & $0^+$  & 2.64  & $10^{4}$    & 0.381   & B & 12.020 \footnotemark[13] \\
 11.914 & 3.772 & $[2^+]$ & 122.4  &   19.7 & 1.77 & $10^{5}$ 
  & $0^+$  & 2.40  & $10^{4}$    & 0.363   & B & 12.250 \footnotemark[13] \\
 12.003 & 3.861 & $[1^-]$ 
  & \multicolumn{2}{c}{$-$} \footnote{state adopted from Ref.~\cite{Berg03}}
  & 4.31 & $10^{5}$ 
  & $1^-$  & 3.84  & $10^{4}$    & 0.354   & A & 12.280 \footnotemark[13] \\
 12.185 & 4.043 & $[3^-]$ & 236.4  &   52.0 & 2.60 & $10^{5}$ 
  & $2^+$  & 9.50  & $10^{4}$    & 0.077   & A & 12.390 \footnotemark[13] \\
 12.474 & 4.332 & $[2^+]$ & 193.8  &   51.6 & 3.89 & $10^{5  }$ 
  & $2^+$  & 6.96  & $10^{4}$    & 0.039   & A & 12.610 \footnotemark[13] \\
 12.665 & 4.523 & $[3^-]$ & 128.8  &   23.5 & 3.45 & $10^{5  }$ 
  & $3^-$  & 6.56  & $10^{4}$    & 0.050   & A & 12.890 \footnotemark[13] \\
 13.010 & 4.868 & $[0^+]$ & 600.9  &  114.5 & 2.16 & $10^{5  }$ 
  & $0^+$  & 5.63  & $10^{4}$    & 0.047   & A & 12.990 \footnotemark[13] \\
\hline
\end{tabular}
%
%\end{center}
\end{table*}

\begin{figure}[htb]
\includegraphics[width=\columnwidth,clip=]{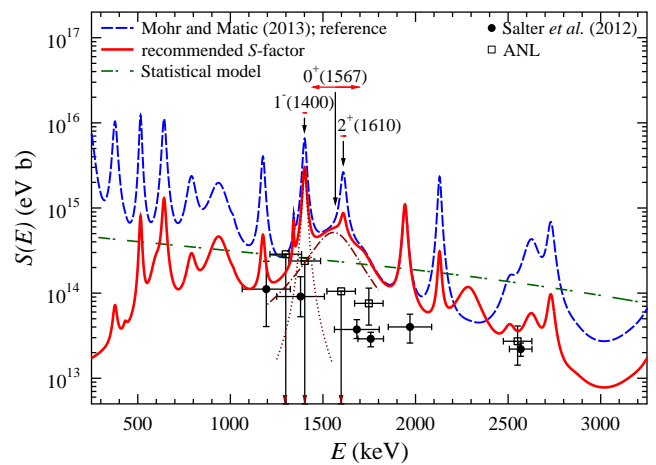}
\caption{
\label{fig:sfact}
(Color online)
Astrophysical \sfact\ of the $^{18}$Ne\rap $^{21}$Na reaction versus
center-of-mass energy $E$, calculated
from the resonance properties listed in Tab.~\ref{tab:str} (thick red
line). For comparison: (dashed blue line) Ref.~\cite{Mohr13}, so-called
reference (for details see \cite{Mohr13});  (dash-dotted
green line) Hauser-Feshbach statistical model, taken from \cite{Rau00}. The
experimental data points shown are derived from the reverse $^{21}$Na\rpa
$^{18}$Ne reaction \cite{Sal12,ANL} and represent the ground state
contribution $^{18}$Ne$_{\rm{g.s.}}$\rap $^{21}$Na$_{\rm{g.s.}}$ only. The
contributions of the two resonances at 1400~keV and 1567~keV are shown with 
dashed and dash-dotted dark-red lines, respectively. The total widths,
$\Gamma$, are indicated by horizontal dark-red arrows. See the text.
}
\end{figure}

The astrophysical S-factor, $S(E)$, of the $^{18}$Ne\rap $^{21}$Na reaction,
calculated using the present resonance strengths that are listed in
Tab.~\ref{tab:str}, is displayed in Fig.~\ref{fig:sfact}. The total 
widths, $\Gamma$, are also adopted from experimental
data (Ref.~\cite{Mohr13} and Tab.~\ref{tab:str}). 
%In those cases where only
%upper limits on $\theta_\alpha^2$ are available, these limits were used. 
For states in category D where reduced widths $\theta_\alpha^2$ are unknown
the numbers in parenthesis in Table \ref{tab:str} were used.
The present S-factor is slightly smaller than the result of Ref.~\cite{Mohr13} in the energy region
1300\,keV $\lesssim E \lesssim$ 2200\,keV, and is significantly smaller at lower
and higher energies. Our smaller S-factor at very low energies results partly
from the new spin assignments and partly from the smaller reduced widths, $\theta_\alpha^2$, and
resonance strengths, $\omega\gamma_{\alpha p}$, for the category D resonances. 
The smaller S-factor at
higher energies, around 2500~keV, results mainly from the new spin assignments of category D
levels. Notice in the figure the broad peak resulting from the $0^+$ resonance at 1567~keV, which will
strongly contribute to the reaction rate (Sec.~\ref{sec:10066}). A similar statement holds
for the $1^-$ resonance at 938~keV, which impacts the reaction rate at
temperatures slightly below $T_9 = 1$.

The Hauser-Feshbach model calculation (dash-dotted green line) is not
able to reproduce the individual resonances in the $^{18}$Ne\rap $^{21}$Na
S-factor. Nevertheless, the general trend of the energy
dependence is approximately reproduced. 
Thus, it can be expected that the statistical model is able to provide the
correct order of magnitude for the reaction rate \Nsv .
As already discussed in Sec.~\ref{sec:basic}, the experimental cross section
of the reverse $^{21}$Na\rpa $^{18}$Ne reaction can be used to derive the ground
state contribution of the $^{18}$Ne$_{\rm{g.s.}}$\rap $^{21}$Na$_{\rm{g.s.}}$ cross
section. Both available data sets (6 data points in \cite{Sal12} and 2 data
points and 3 upper limits in \cite{ANL}) are in reasonable agreement with
each other. 
As expected, the experimental ground state contribution is lower than
the present total S-factor.

%%%%%%%%%%%%%%%%%%%%%%%%%%%%%%%%%%%%%%%%%

\subsection{The $E^\ast = 10066$\,keV level in $^{22}$Ne}
\label{sec:10066}
The level at $E^\ast = 10066$\,keV in $^{22}$Ne,
whose mirror in $^{22}$Mg corresponds to a strong
resonance at 1567\,keV in the
$^{18}$Ne\rap $^{21}$Na reaction, requires special attention. This state is tentatively assigned as
$J^\pi = (0^+)$ in Ref.~\cite{ENSDF}, but $J^\pi = 1^-$ or even $1^+$
have also been suggested. In the present work we adopt $E^\ast = 10052$\,keV and $J^\pi = 0^+$,
based on the following arguments. 

First, the state must have natural parity because it has been observed in \al -transfer \cite{Gie94} and \al -capture
\cite{Dab03} studies. In the electron scattering work of Ref.~\cite{Mar74} it is stated
that ``$\ldots$ two states (10.08 and 12.56\,MeV) are compatible with
either $J^\pi = 1^+$ or $1^-$.'' However, a $2^+$ assignment for the 10066\,keV
state was only excluded on purely theoretical grounds. 
A comparison of excitation energies from the electron scattering experiment
\cite{Mar74} and the 
ENSDF database \cite{ENSDF} shows an offset by about 50\,keV for neighboring
states, implying a corrected excitation energy of $E^\ast \approx
10.080\,{\rm{MeV}} + 0.05\,{\rm{MeV}} = 10.130$\,MeV. This value is in excellent agreement
with the location of a $J^\pi = 2^+$ state at $E^\ast = 10.137$\,MeV in $^{22}$Ne \cite{ENSDF}. Assuming an assignment of $J^\pi = 2^+$ results in a reduced 
transition strength of $B(E2)\!\!\downarrow \, = 8^{+1.6}_{-0.9}\,e^2$fm$^4$, which was excluded in Ref.~\cite{Mar74} only because shell model
calculations at that time predicted $B(E2)\!\!\downarrow \, \le 3\,e^2$fm$^4$ (or
$\le 0.8$\,W.u.). However, modern calculations \cite{Lev13} result in larger transition strengths. Thus, it is likely that the $2^+$ state at
$E^\ast = 10137$\,keV in $^{22}$Ne has been detected by Ref.~\cite{Mar74}, and that the $J^\pi$
assignment in Ref.~\cite{Mar74} is incorrect.

A photon scattering experiment \cite{Berg79} did not observe
any peak near $E^\ast = 10066$\,keV, consistent with a $J^\pi = 0^+$
assignment. In principle, the $2^+$ state at $E^\ast
= 10137$\,keV could have been observed in the photon scattering experiment. However, 
the reduced transition strength measured in the electron scattering experiment
\cite{Mar74} results in 2.2\,W.u.\ for the ground state transition, which is most likely below
the detection limit in the photon scattering measurement \cite{Berg79}. 

The $J^\pi = 0^+$ assignment for the state at 10066\,keV is also preferred by
($^6$Li,$d$) transfer data \cite{Gie94}, although $J^\pi
= 1^-$ cannot be ruled out. Furthermore, several theoretical calculations
predict a $J^\pi = 0^+$ state with a large $\alpha$-particle reduced width, $\theta_\alpha^2$, near
$E^\ast \approx 10$\,MeV \cite{Lev13,Kim07,Des88}, consistent with
the large reduced widths found both in \al -capture and \al -transfer. 

Consequently, we assign here $J^\pi = 0^+$ to this level, and
we adopt a slightly lower excitation energy of $E^\ast = 10052$\,keV, based
on \al -capture and \al -transfer data \cite{Dab03,Gie94} and disregarding the value of
$E^\ast = 10080$\,keV from electron scattering \cite{Mar74}.

The measured strength of the corresponding resonance in
$^{18}$O($\alpha$,$\gamma$)$^{22}$Ne amounts to $\omega
\gamma_{\alpha\gamma}^{exp}$ = $0.24 \pm 0.08$\,$\mu$eV \cite{Dab03}. The
uncertainty is relatively large and yields a value of $\theta_\alpha^2$ =
$0.27 \pm 0.09$ for the dimensionless reduced $\alpha$-particle width. It
should be noted that the \al -capture experiment \cite{Dab03} applied a
coincidence technique, where the resonance strength $\omega
\gamma_{\alpha\gamma}^{exp}$ was derived using the measured yield of the $2^+
\rightarrow 0^+$ transition from the first excited state in $^{22}$Ne to the
ground state. This measured {\it partial} resonance strength has been
increased by Ref.~\cite{Dab03} to an estimated total strength of $\omega
\gamma_{\alpha\gamma}^{set} = 0.48 \pm 0.16$\,$\mu$eV, by taking a typical
direct ground state branching ratio of $\approx$50\,\% for $1^-$ states in
$^{22}$Ne into account. However, with our newly adopted $J^\pi = 0^+$
assignment the direct $0^+ \rightarrow 0^+$ ground state branching ratio
becomes very small because $0^+ \rightarrow 0^+$ transitions cannot proceed
via a direct $\gamma$-ray transition. 
Since cascade transitions bypassing the first excited $2^+$ state have
typically very small contributions, only a minor correction should be applied
to the measured partial resonance strength from Ref.~\cite{Dab03}.

From the relative spectroscopic factor of 0.15 reported in the \al -transfer
study of Ref.~\cite{Gie94}, a value of $\theta_\alpha^2 = 0.64$ can be derived
using the procedure outlined above in Sect.~\ref{sec:mirror}. 
Together with
the result from the \al -capture measurement quoted above, we adopt an average
value of $\theta_\alpha^2 = 0.46$ for this state. The resulting resonance
strength in the $^{18}$Ne\rap $^{21}$Na reaction amounts to $\omega
\gamma_{\alpha p}$ = $65$\,eV, with an estimated uncertainty of a factor of
two.

%%%%%%%%%%%%%%%%%%%%%%%%%%%%%%%%%%%%%%%%%%%%%%%%%%%%%

\section{Monte Carlo-based reaction rates}
\label{sec:MC}
The $^{18}$Ne\rap $^{21}$Na reaction rate can be calculated directly from
Eq.~(\ref{eq:rate}) by adopting resonance energies, $E$, from
Ref.~\cite{Matic09} and resonance strengths, $\omega\gamma_{\alpha p}$, from
Table \ref{tab:str}. 
More than 30 resonances enter into the sum of Eq.~(\ref{eq:rate}). Thus
the uncertainties of more than 60 parameters 
(resonance energies and resonance strengths)
must be considered simultaneously for the total
reaction rate. Additional uncertainties arise for some resonances from
ambiguous spin-parity assignments. The latter uncertainties could not reliably
be taken into account in previous work \cite{Matic09,Mohr13}.  

Here we present for the first time an evaluation of the experimental
$^{18}$Ne\rap $^{21}$Na reaction rate using the Monte Carlo method introduced
by Refs.~\cite{Long10,Ili10a,Ili10b,Ili10c}. The method of
\cite{Long10,Ili10a,Ili10b,Ili10c} had to be extended for a proper treatment
of uncertain $J^\pi$ assignments which will be discussed in detail later.
Briefly summarizing \cite{Long10,Ili10a,Ili10b,Ili10c}, for each
input parameter (i.e., in the present case for each resonance energy, $E$, and
each resonance strength, $\omega \gamma$) a probability density function is obtained, based on the experimental mean value and the uncertainty. We assumed a Gaussian probability density
function for resonance energies, and a lognormal one for resonance strengths.
For those resonances with unknown
    strengths, $\alpha$-particle partial widths were sampled
    according to a Porter-Thomas distribution with an absolute upper limit
    determined from Eq.~(\ref{eq:wigner}).
These choices are justified in Refs.~\cite{Long10,Pog13}. 
These input probability densities are sampled many times and each time a
reaction rate sample is obtained. The combined ensemble of samples can then be
used to define statistically meaningful recommended reaction rates and
uncertainties that correspond to a desired coverage probability. In this work,
we adopt the 50th percentile of the cumulative rate distribution as the
recommended rate, \Nsv $_{\mathrm{rec}}$, and the 16th and 84th percentiles as
the low rate, \Nsv $_{\rm{low}}$, and the high rate, \Nsv 
$_{\rm{high}}$, respectively (for a coverage probability of 68\%). The
above percentiles are taken in analogy to the 1$\sigma$ uncertainty of the
usual Gaussian distribution. It is also possible to provide a 95\% coverage
interval, analog to the 2$\sigma$ uncertainty. However, it should be kept in
mind that the lognormal distribution of the resonance strengths leads to an
approximately lognormal distribution of the reaction rate, and thus the low
and the high rate are not symmetric around the recommended median rate. Some
numrical values will be given later for the temperature of $T_9 = 1$. Further
technical details and probability density distributions for the reaction rate
\Nsv\ are given in \cite{Mohr14NIC}.

In a first calculation all spins and parities $J^\pi$ in Table \ref{tab:str}
were used. However, because some $J^\pi$ assignments are only tentative, this
calculation underestimates the uncertainty of the reaction rate
\Nsv . Therefore, in a second calculation we took into account these
uncertain $J^\pi$ assignments, and an improved estimate of the resulting
uncertainty of \Nsv\ could be derived.

\begin{figure}[htb]
\includegraphics[width=\columnwidth,clip=]{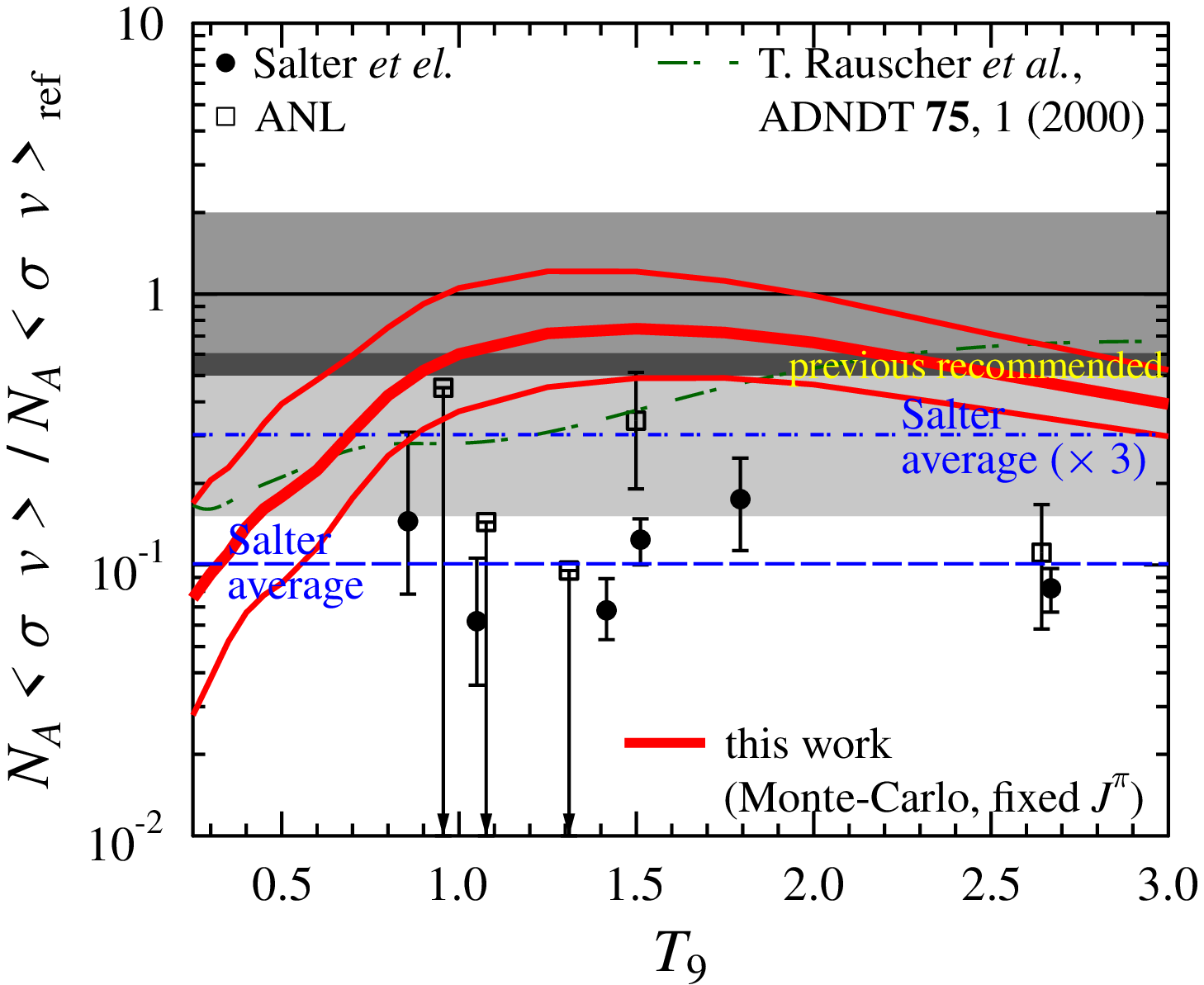}
\caption{
\label{fig:rate_fix}
(Color online)
Ratio between the reaction rate \Nsv\ from different studies
normalized to the reference rate \Nsv $_{\rm{ref}}$ from
\cite{Mohr13}. The conversion of the experimental data from the reverse
$^{21}$Na\rpa $^{18}$Ne reaction to the shown \Nsv\ data points is explained
in \cite{Mohr13}. The line ``Salter average'' marks the average result from
\cite{Sal12} (ground state contribution only), ``Salter average ($\times 3$)''
corrects for contributions of excited states (for details see
\cite{Mohr13,Sal12}).
The new result (thick red line) and its uncertainties
(thin red lines) are close to the previous recommendation which is 0.55 \Nsv
$_{\rm{ref}}$ \cite{Mohr13} (dark grey bar). A theoretical prediction in the
statistical model is also shown (green dash-dotted) \cite{Rau00}. Further
discussion see text. 
}
\end{figure}

The result of the initial Monte-Carlo calculation with the fixed $J^\pi$
assignments in Table \ref{tab:str} is shown in Fig.~\ref{fig:rate_fix}. It is
close to the previous recommendation of 0.55 \Nsv $_{\rm{ref}}$ \cite{Mohr13}
for temperatures $1 \lesssim T_9 \lesssim 2$ and slightly lower in the low
temperature range of $0.5 \lesssim T_9 \lesssim 1$. The lower rate results
from the unnatural parity of the state at $E^\ast = 8.385$\,MeV and the smaller
reduced widths $\theta_\alpha^2$ of several category D states around $E^\ast =
8.5 - 9$\,MeV.

The method outlined in \cite{Long10,Ili10a,Ili10b,Ili10c} assumed an unique
$J^\pi$ assignment for the random sampling over nuclear physics input
parameters (resonance energies, strengths, S-factors, partial widths,
etc.). Here we extend the method by allowing for ambiguous $J^\pi$
assignments. If the $J^\pi$ value of a given level is not known unambiguously,
we randomly sample over possible $J^\pi$ according to a discrete probability
density. The probabilities assigned to each $J^\pi$ value have to be chosen
according to the best knowledge available (guided by experimental or
theoretical information). If $J^\pi$ has been restricted to a range, and no
other information is available, then the probabilities for each $J^\pi$ value
should be the same.

For the $^{18}$Ne\rap $^{21}$Na reaction under study, the $J^\pi$ assignments
of Table \ref{tab:str} were varied in the following way. The tentative
assignment in Table \ref{tab:str} was assumed with a 50\,\%
probability. (Arguments for these tentative assignments are mainly taken from
\cite{Matic09}.) The
remaining 50\,\% were distributed uniformly among other possible assignments,
e.g.\ taken from Table II of \cite{Chae09} (``$\chi^2$ values for possible $l$
transfers'') or from uncertain assignments in the mirror nucleus
\cite{Gie94}. If there is no restriction on $J^\pi$ from resonant elastic
scattering or from \al\ transfer reactions, the remaining 50\,\% were
distributed uniformly among all natural parity states from $0^+$ to
$4^+$. The chosen values are listed in Table \ref{tab:spin_var}. In cases with
a well-defined $J^\pi$ assignment in the $^{22}$Ne mirror nucleus (e.g., from
\al -transfer in \cite{Gie94}), no variation of $J^\pi$ in $^{22}$Mg was
allowed, and the adopted $J^\pi$ of Table \ref{tab:str} was used. This is
justified by a cancellation effect which will be discussed later (see
Sect.~\ref{sec:canc}).
\begin{table}[tbh]
\caption{\label{tab:spin_var}
Variable $J^\pi$ assignments for the improved estimate of the uncertainty of
the reaction rate \Nsv\ of the $^{18}$Ne\rap $^{21}$Na
reaction. A probability of $p = 1/2$ is assumed for the $J^\pi$ assignments in
Table \ref{tab:str} (marked in bold). The remaining 50\,\% are distributed
among other possible $J^\pi$. The given numbers $p(J^\pi)$ are used as
discrete probability densities for Monte-Carlo sampling of the reaction rate
\Nsv . Further details see text.
}
%\begin{center}
\begin{tabular}{cccc}
\hline
\multicolumn{1}{c}{$E^\ast(^{22}{\rm{Mg)}}$ (MeV)}
& \multicolumn{1}{c}{$E$ (MeV)}
& \multicolumn{1}{c}{$J^\pi$}
& \multicolumn{1}{c}{$p(J^\pi)$} \\
\hline
  9.318 & 1.176 & $0^+$ & $1/8$ \\
        &       & $1^-$ & $1/8$ \\
        &       & $\mathbf{2^+}$ & $\mathbf{1/2}$ \\
        &       & $3^-$ & $1/8$ \\
        &       & $4^+$ & $1/8$ \\
  9.482 & 1.340 & $2^+$ & $1/4$ \\
        &       & $\mathbf{3^-}$ & $\mathbf{3/4}$ \\
  9.752 & 1.610 & $1^-$ & $1/2$ \\
        &       & $\mathbf{2^+}$ & $\mathbf{1/2}$ \\
  9.860 & 1.718 & $\mathbf{0^+}$ & $\mathbf{1/2}$ \\
        &       & $1^-$ & $1/4$ \\
        &       & $2^+$ & $1/4$ \\
 10.085 & 1.943 & $0^+$ & $1/4$ \\
        &       & $1^-$ & $1/4$ \\
        &       & $\mathbf{2^+}$ & $\mathbf{1/2}$ \\
 10.651 & 2.509 & $0^+$ & $1/8$ \\
        &       & $1^-$ & $1/8$ \\
        &       & $2^+$ & $1/8$ \\
        &       & $\mathbf{3^-}$ & $\mathbf{1/2}$ \\
        &       & $4^+$ & $1/8$ \\
 10.768 & 2.626 & $0^+$ & $1/8$ \\
        &       & $1^-$ & $1/8$ \\
        &       & $\mathbf{2^+}$ & $\mathbf{1/2}$ \\
        &       & $3^-$ & $1/8$ \\
        &       & $4^+$ & $1/8$ \\
 10.873 & 2.731 & $\mathbf{0^+}$ & $\mathbf{1/2}$ \\
        &       & $1^-$ & $1/8$ \\
        &       & $2^+$ & $1/8$ \\
        &       & $3^-$ & $1/8$ \\
        &       & $4^+$ & $1/8$ \\
 11.001 & 2.859 & $0^+$ & $1/8$ \\
        &       & $1^-$ & $1/8$ \\
        &       & $2^+$ & $1/8$ \\
        &       & $3^-$ & $1/8$ \\
        &       & $\mathbf{4^+}$ & $\mathbf{1/2}$ \\
 11.315 & 3.173 & $0^+$ & $1/8$ \\
        &       & $1^-$ & $1/8$ \\
        &       & $2^+$ & $1/8$ \\
        &       & $3^-$ & $1/8$ \\
        &       & $\mathbf{4^+}$ & $\mathbf{1/2}$ \\
 11.595 & 3.453 & $3^-$ & $1/3$ \\
        &       & $\mathbf{4^+}$ & $\mathbf{2/3}$ \\
\hline
\end{tabular}
%
%\end{center}
\end{table}

Although carefully chosen, it is obvious that the above assumptions on the
chosen probabilities of $J^\pi$ are somewhat
arbitrary. Nevertheless, the derived \Nsv\ and in particular its uncertainty
should be more realistic than all previous estimates where in most cases only
fixed $J^\pi$ assignments were considered (i.e., neglecting any uncertain
$J^\pi$). As the other extreme, a fully random $J^\pi$ assignment
was used in \cite{Matic09} to estimate an upper limit of the uncertainty of
\Nsv\ from the $J^\pi$ assignments; here an uncertainty of about a factor of
10 was found.

\begin{figure}[htb]
\includegraphics[width=\columnwidth,clip=]{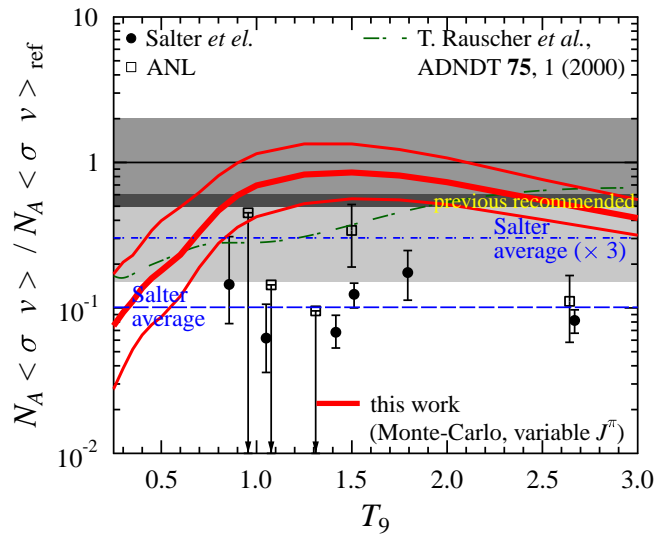}
\caption{
\label{fig:rate_var}
(Color online)
Same as Fig.~\ref{fig:rate_fix}, but with the additional uncertainty from
the spin assignments $J^\pi$ in Table \ref{tab:spin_var}. The new reaction rate
\Nsv\ using variable $J^\pi$ assignments is slightly higher than the first
calculation with fixed $J^\pi$ in Fig.~\ref{fig:rate_fix} but remains close to
the previous recommendation which is 0.55 \Nsv $_{\rm{ref}}$ \cite{Mohr13}
(dark grey bar). 
}
\end{figure}

Using the larger parameter space with variable $J^\pi$ assignments from
Table \ref{tab:spin_var} in our Monte-Carlo sampling, we find that the
reaction rate \Nsv\ remained within 
about 20\,\% of the result with fixed $J^\pi$ assignments. The reason for
these minor changes is that variable $J^\pi$ assignments have to be taken into
account mainly for category D states (without spectroscopic information) which
have relatively small resonance strengths because of their small reduced widths
$\theta_\alpha^2$. The result is shown in Fig.~\ref{fig:rate_var}, and
numerical values are listed in Table \ref{tab:rate}. The overall
uncertainty is slightly above a factor of two at low temperatures ($T_9
\approx 0.5$), and reduces to about a factor of 1.5 for $1 \lesssim T_9
\lesssim 3$. The uncertainty in the total rate can be smaller than the
uncertainty in individual resonance parameters 
because the rate calculation averages over the contributing resonances.
For simple use in astrophysical calculations, a fit to the reaction rate 
\Nsv\ of the $^{18}$Ne\rap $^{21}$Na reaction is provided in the Appendix
\ref{sec:rate_fit}. 
\begin{table}[tbh]
\caption{\label{tab:rate}
Recommended reaction rate \Nsv $_{\rm{rec}}$ of the
$^{18}$Ne\rap $^{21}$Na reaction (in cm$^3$\,s$^{-1}$\,mol$^{-1}$) from
Monte-Carlo sampling. The underlying input parameters for the Monte-Carlo
approach are based on experimental information from various sources (for
details see text). 
}
%\begin{center}
\begin{tabular}{p{0.8cm}ccc}
\hline
\multicolumn{1}{c}{$T_9$}
& \multicolumn{1}{c}{low}
& \multicolumn{1}{c}{rec}
& \multicolumn{1}{c}{high} \\
\hline
0.1 & 2.00$\times$10$^{-27}$ & 5.94$\times$10$^{-27}$ &
    1.76$\times$10$^{-26}$ \\
0.2 & 4.36$\times$10$^{-17}$ & 1.12$\times$10$^{-16}$ &
    2.74$\times$10$^{-16}$ \\
0.3 & 2.04$\times$10$^{-12}$ & 5.00$\times$10$^{-12}$ &
    1.11$\times$10$^{-11}$ \\
0.4 & 1.33$\times$10$^{-09}$ & 2.78$\times$10$^{-09}$ &
    5.60$\times$10$^{-09}$ \\
0.5 & 1.13$\times$10$^{-07}$ & 2.35$\times$10$^{-07}$ &
    5.12$\times$10$^{-07}$ \\
0.6 & 3.69$\times$10$^{-06}$ & 7.09$\times$10$^{-06}$ &
    1.51$\times$10$^{-05}$ \\
0.7 & 7.10$\times$10$^{-05}$ & 1.25$\times$10$^{-04}$ &
    2.36$\times$10$^{-04}$ \\
0.8 & 8.54$\times$10$^{-04}$ & 1.44$\times$10$^{-03}$ &
    2.49$\times$10$^{-03}$ \\
0.9 & 6.67$\times$10$^{-03}$ & 1.11$\times$10$^{-02}$ &
    1.87$\times$10$^{-02}$ \\
1.0 & 3.72$\times$10$^{-02}$ & 6.09$\times$10$^{-02}$ &
    1.02$\times$10$^{-01}$ \\
1.1 & 1.57$\times$10$^{-01}$ & 2.54$\times$10$^{-01}$ &
    4.22$\times$10$^{-01}$ \\
1.2 & 5.28$\times$10$^{-01}$ & 8.48$\times$10$^{-01}$ &
    1.39$\times$10$^{+00}$ \\
1.3 & 1.49$\times$10$^{+00}$ & 2.37$\times$10$^{+00}$ &
    3.86$\times$10$^{+00}$ \\
1.4 & 3.70$\times$10$^{+00}$ & 5.76$\times$10$^{+00}$ &
    9.23$\times$10$^{+00}$ \\
1.5 & 8.16$\times$10$^{+00}$ & 1.25$\times$10$^{+01}$ &
    1.97$\times$10$^{+01}$ \\
1.6 & 1.64$\times$10$^{+01}$ & 2.47$\times$10$^{+01}$ &
    3.84$\times$10$^{+01}$ \\
1.7 & 3.06$\times$10$^{+01}$ & 4.54$\times$10$^{+01}$ &
    6.94$\times$10$^{+01}$ \\
1.8 & 5.34$\times$10$^{+01}$ & 7.83$\times$10$^{+01}$ &
    1.18$\times$10$^{+02}$ \\
1.9 & 8.84$\times$10$^{+01}$ & 1.28$\times$10$^{+02}$ &
    1.89$\times$10$^{+02}$ \\
2.0 & 1.40$\times$10$^{+02}$ & 2.01$\times$10$^{+02}$ &
    2.93$\times$10$^{+02}$ \\
2.1 & 2.13$\times$10$^{+02}$ & 3.03$\times$10$^{+02}$ &
    4.37$\times$10$^{+02}$ \\
2.2 & 3.13$\times$10$^{+02}$ & 4.42$\times$10$^{+02}$ &
    6.30$\times$10$^{+02}$ \\
2.3 & 4.49$\times$10$^{+02}$ & 6.27$\times$10$^{+02}$ &
    8.84$\times$10$^{+02}$ \\
2.4 & 6.31$\times$10$^{+02}$ & 8.71$\times$10$^{+02}$ &
    1.21$\times$10$^{+03}$ \\
2.5 & 8.69$\times$10$^{+02}$ & 1.19$\times$10$^{+03}$ &
    1.64$\times$10$^{+03}$ \\
2.6 & 1.17$\times$10$^{+03}$ & 1.59$\times$10$^{+03}$ &
    2.18$\times$10$^{+03}$ \\
2.7 & 1.57$\times$10$^{+03}$ & 2.10$\times$10$^{+03}$ &
    2.86$\times$10$^{+03}$ \\
2.8 & 2.06$\times$10$^{+03}$ & 2.75$\times$10$^{+03}$ &
    3.71$\times$10$^{+03}$ \\
2.9 & 2.69$\times$10$^{+03}$ & 3.55$\times$10$^{+03}$ &
    4.75$\times$10$^{+03}$ \\
3.0 & 3.44$\times$10$^{+03}$ & 4.55$\times$10$^{+03}$ &
    6.06$\times$10$^{+03}$ \\
\hline
\end{tabular}
%
%\end{center}
\end{table}

For interpretation of the rate \Nsv\ and the given uncertainties in Table
\ref{tab:rate}, we discuss in detail the result for the temperature $T_9 =
1.0$. All \Nsv\ in the following discussion are given in $10^{-3}$\,
cm$^3$\,s$^{-1}$\,mol$^{-1}$ (without explicitly repeating this unit).
Our recommended result is 60.9 which is the median of the 10,000 Monte-Carlo
samples which were calculated. The 16th percentile gives the recommended lower
rate of 37.2; i.e., 1,600 of the 10,000 Monte-Carlo samples provided a rate
below 37.2. Similar to the recommended lower rate, the recommended upper rate
of 102 is defined at 84th percentile. The uncertainty of the new recommended
rate of 60.9 is a factor of 1.67, covering 68\,\% of the probability density
distribution and reflecting the approximately lognormal
distribution of the calculated reaction rates. A wider uncertainty band with
95\,\% coverage can be taken from the 2.3rd and
97.7th percentile of the Monte-Carlo samples, leading to a wider range for
\Nsv\ between 23 and 184.

\begin{figure}[htb]
\includegraphics[width=\columnwidth,clip=]{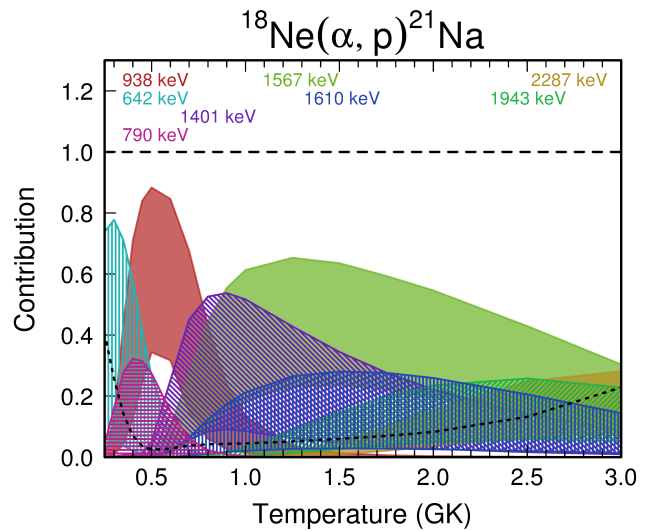}
\caption{
\label{fig:contrib}
(Color online)
Contributions of individual resonances to the reaction rate \Nsv . Only
few resonances contribute significantly to \Nsv\ in the most important
temperature range around $T_9 = 0.5 - 2$. 
The black dashed line corresponds to the total contribution
  of all other resonances not explicitly displayed in the plot.
}
\end{figure}

The Monte Carlo method also provides the fractional contributions of
individual resonances to the total rate. 
On a sample-by-sample basis, the fractional contributions of all
    resonances can be computed to obtain an ensemble of
    contributions over the full Monte Carlo calculation. 
Similar to the
    definition of low and high reaction rates via percentiles, for these
    ensembles of fractional contributions low and high contributions are given
    at the 16th and 84th 
    percentiles.
This method ensures that individual contributions and
    their uncertainties do not exceed unity or become negative.
These contributions are displayed in Fig.~\ref{fig:contrib}. 
It is
apparent that the total rate is dominated by few strong resonances. At
typical break-out temperatures slightly below $T_9 = 1$, the $1^-$
resonance at 938\,keV (shown in red) contributes more than 50\,\% to
the total rate. At slightly higher temperatures around $T_9 = 1 - 2$,
the $1^-$ resonance at 1401\,keV, the broad $0^+$ resonance at
1567\,keV, and at moderate extent the $(1^-,2^+)$ resonance at 1610\,keV
dominate the total rate. 

\section{Discussion}
\label{sec:disc}
\subsection{Comparison to previous work}
\label{sec:comp}
Our recommended rate is
slightly higher than the previous rate \cite{Mohr13} at intermediate temperatures ($T_9 = 1 - 2$), slightly
lower at high temperatures ($T_9 \gg 2$), and significantly lower for very low
temperatures ($T_9 < 0.5$). These differences are a direct consequence of the changes in the astrophysical \sfact\ (Fig.~\ref{fig:sfact}) that were discussed in Sec.~\ref{sec:mirror}.

The reaction rate of Ref.~\cite{Matic09} is much higher than the
present result because the former was obtained by using some huge strengths
measured in Ref.~\cite{Gro02} that are in contradiction to the experimental 
data for the reverse reaction \cite{Sal12,ANL}, as pointed out by Ref.~\cite{Mohr13}.

The reaction rate of Salter {\it et al.}\ \cite{Sal12} is much lower than the
present result since it was directly derived from the experimental reverse
reaction data. As noted above, this procedure provides the ground state
contribution only, which obviously must be smaller than the total reaction
rate.

The reaction rate obtained in the recent study of $^{21}$Na\rpp $^{21}$Na
resonant elastic scattering by Zhang {\it et al.}\ \cite{Zhang14} is close to
the earlier recommendation of Ref.~\cite{Mohr13}. In general, Zhang {\it et
  al.}\ adopt the resonance strengths from Ref.~\cite{Mohr13} because the new
$J^\pi$ assignments of Ref.~\cite{Zhang14} confirm for most levels 
earlier tentative assignments adopted in Refs.~\cite{Matic09,Mohr13}. New
spin-parity assignments have been made only for a few states, and in
particular the resonance at 243\,keV has been excluded because of its
unnatural parity. This leads to slightly reduced \Nsv\ at low temperatures in
\cite{Zhang14} compared to Ref.~\cite{Mohr13}.

\subsection{An interesting cancellation effect}
\label{sec:canc}
The remaining influence of uncertain mirror assignments on \Nsv\ is
surprisingly small. This is mainly based on an interesting cancellation
effect. The following discussion extends a similar idea which was already
presented by Iliadis {\it et al.}\ \cite{Ili99} for a fictitious resonance in
the $^{35}$Ar\rpg $^{36}$K reaction. Let us consider here the $0^+$ resonance
at 1567\,keV as an example.

Matic {\it et al.}\ \cite{Matic09} have assigned the $0^+$ state at $E^\ast =
10052$\,keV in $^{22}$Ne as the mirror state. For this mirror state a large
reduced width of $\theta_\alpha^2 = 0.46$ has been found in the $^{18}$O\rag
$^{22}$Ne and $^{18}$O($^6$Li,$d$)$^{22}$Ne reactions
\cite{Dab03,Vog90,Tra78,Gie94}; thus, the \al -cluster properties of this
state in $^{22}$Ne are well-established from experiment, and also theory
suggests such a state \cite{Lev13,Kim07,Des88}. Based on mirror
symmetry, a state with a similar \al -cluster wave function should exist also
in $^{22}$Mg around $E^\ast \approx 10$\,MeV. 

Let us now first assume that the mirror assignment in \cite{Matic09} is
correct. Then this state appears as a resonance in the $^{18}$Ne\rap
$^{21}$Na reaction at $E = 1567$\,keV with a resonance strength of $\omega
\gamma = 65$\,eV. 
It contributes to the rate with \Nsv\ $= 5.1$\,cm$^3$\,s$^{-1}$\,mol$^{-1}$
e.g.\ at $T_9 = 1.5$. 

Let us next consider what 
happens if this mirror assignment is incorrect. If the \al -strength with
$\theta_\alpha^2 = 0.46$ is located about 250\,keV lower (higher), then the
resonance strength decreases (increases) by about one order of
magnitude to $\omega \gamma = 6.2$\,eV (422\,eV). Nevertheless, the changes in
\Nsv\ at $T_9 = 1.5$ remain within less than a factor of two with \Nsv\ $=
3.2$\,cm$^3$\,s$^{-1}$\,mol$^{-1}$ (4.6\,cm$^3$\,s$^{-1}$\,mol$^{-1}$) for
the lower (higher) energy. 

\begin{figure}[htb]
\includegraphics[width=\columnwidth,clip=]{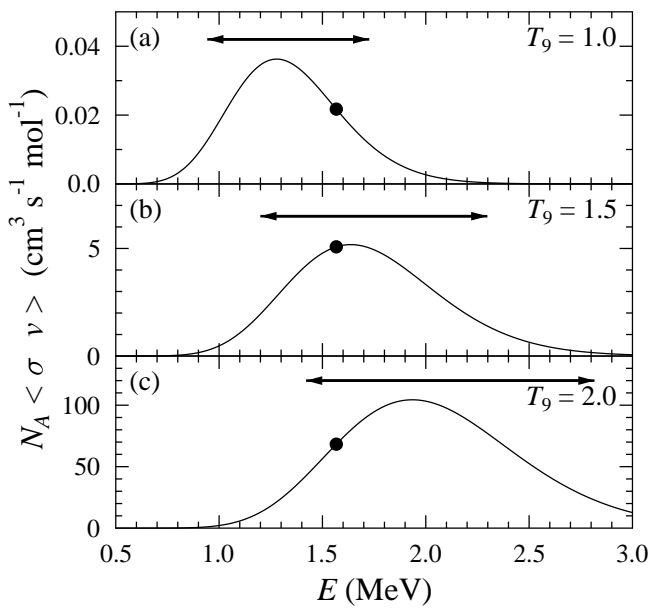}
\caption{
\label{fig:gamow}
Reaction rate \Nsv\ for a $0^+$ resonance with $\theta_\alpha^2 = 0.46$
in $^{22}$Mg as a function of resonance energy $E$ for three temperatures
$T_9 = 1.0$ (a), 1.5 (b), and 2.0 (c) (from top to bottom). The data point
indicates $E = 1567$\,keV which has been assigned as mirror of the $0^+$ state
in $^{22}$Ne at $E^\ast = 10052$\,keV. The arrows indicate the position of the
usual Gamow window. For further discussion see text.
}
\end{figure}

The explanation for this mild dependence can be
read from Eq.~(\ref{eq:rate}). The reaction rate \Nsv\ scales exponentially
with the resonance energy $E$ in the factor $\exp{(-11.605E/T_9)}$, and it scales
linearly with the resonance strength $\omega \gamma \approx (2J+1)
\Gamma_\alpha = (2J+1) \, \theta_\alpha^2 \, \Gamma_\alpha^{s.p.}$. However,
$\Gamma_\alpha^{s.p.}$ scales exponentially with the resonance energy. Thus,
for a given $\theta_\alpha^2$ (determined in the mirror nucleus) 
we find that \Nsv\ is the product of an exponentially rising resonance
strength and an exponentially decreasing factor $\exp{(-11.605E/T_9)}$ (the
latter factor results from the Maxwell-Boltzmann distribution). This product
shows a maximum very similar to the conventional Gamow window. For the given
example this is illustrated in Fig.~\ref{fig:gamow}; the above discussed
temperature of $T_9 = 1.5$ is shown in the middle part.

At $T_9 = 1.5$ the resonance at 1567\,keV is located almost in the center of
the usual Gamow window which is given by the most effective energy $E_0 =
0.122 \, (Z_1^2 Z_2^2 A_{\rm{red}} T_9^2)^{1/3}$\,MeV and the $1/e$ width $\Delta
= 0.237 \, (Z_1^2 Z_2^2 A_{\rm{red}} T_9^5)^{1/6}$\,MeV (see
Refs.~\cite{Ili07,Rol88}). Thus, the reaction rate \Nsv\ is close to its
maximum for 
$E = 1567$\,keV, and \Nsv\ decreases slightly for lower and higher resonance
energies $E$. 

At lower temperatures, e.g.\ $T_9 = 1.0$ (Fig.~\ref{fig:gamow}, upper part),
the most effective energy $E_0$ is lower, and the resonance at $E = 1567$\,keV
is located on the high-energy side of the Gamow window. For $E = 1567$\,keV a
reaction rate \Nsv\ $ = 0.022$\,cm$^3$\,s$^{-1}$\,mol$^{-1}$ is found. If the
resonance is lowered in energy by 250\,keV, now the rate increases to
\Nsv\ $ = 0.036$\,cm$^3$\,s$^{-1}$\,mol$^{-1}$ because this lower resonance
energy is located in the center of the Gamow window. Increasing the resonance
energy by 250\,keV leads to a further reduced rate of \Nsv\ $ =
0.008$\,cm$^3$\,s$^{-1}$\,mol$^{-1}$.

At higher temperatures, e.g.\ $T_9 = 2.0$ (Fig.~\ref{fig:gamow}, lower part),
the most effective energy $E_0$ is higher, and now $E = 1567$\,keV leads to
\Nsv\ $ = 68$\,cm$^3$\,s$^{-1}$\,mol$^{-1}$. For the resonance with 250\,keV
lowered energy the rate \Nsv\ $ = 27$\,cm$^3$\,s$^{-1}$\,mol$^{-1}$ is
lower, and for the resonance with 250\,keV increased energy the rate
\Nsv\ $ = 100$\,cm$^3$\,s$^{-1}$\,mol$^{-1}$ is higher.

In conclusion this means that the experimentally confirmed \al -cluster state
in $^{22}$Ne at $E^\ast = 10052$\,keV with $\theta_\alpha^2 = 0.46$ leads to a
relatively well-constrained reaction rate \Nsv\ of the $^{18}$Ne\rap $^{21}$Na
reaction for temperatures $T_9 = 1 - 2$ as long as ($i$) the principle of
mirror symmetry is accepted and ($ii$) the mirror state in $^{22}$Mg is
located anywhere within a 500\,keV broad window around the currently accepted
mirror assignment at $E^\ast = 9709$\,keV ($E = 1567$\,keV).

The above argument can be generalized. It is very helpful to determine reduced
widths $\theta_\alpha^2$ (e.g.\ from the mirror system) for resonances which
are located in the classical Gamow window because the reduced width
$\theta_\alpha^2$ essentially constrains the reaction rate \Nsv\ even in cases
when the mirror assignment is unclear or only tentative.

\section{Summary and conclusions}
\label{sec:summ}
The stellar reaction rate \Nsv\ of the $^{18}$Ne\rap $^{21}$Na reaction is
composed of the contributions of 32 resonances. The Monte-Carlo method
has been applied to calculate \Nsv\ from experimental resonance energies
\cite{Matic09} and resonance strengths which were calculated using reduced
widths $\theta_\alpha^2$ either from the mirror nucleus
\cite{Tra78,Vog90,Gie94,Dab03,Gol04} or from a Porter-Thomas distribution found
in a recent systematic study \cite{Pog13}. 
Furthermore, uncertainties from tentative $J^\pi$ assignments were taken into
account by Monte-Carlo sampling of discrete probability distributions.
The recommended result for \Nsv\ is close to
previous recommendations \cite{Mohr13,Zhang14}. Nevertheless, there is
significant progress compared to these previous recommendations: In 
\cite{Mohr13,Zhang14} the recommended rate was composed as a compromise
between a relatively high \Nsv\ from resonance energies and calculated
resonance strengths and a relatively low \Nsv\ from the reverse reaction
data. Now the newly determined resonance strengths are somewhat lower,
thus bringing both approaches for the determination of \Nsv\ in better
agreement and increasing the reliability of the result.

In addition, the Monte-Carlo formalism allows an improved determination of
uncertainties which takes into account the uncertainties of all ingredients in
a consistent way. It turns out that the uncertainty of \Nsv\ is essentially
given by the uncertainties of the resonance strengths whereas the resonance
energies of \cite{Matic09} are sufficiently precise. The uncertainties of the
resonance strengths were estimated according to the available information on
the reduced \al -width $\theta_\alpha^2$ in the $^{22}$Ne mirror nucleus. As
final result we find that \Nsv\ (given as the median of the Monte-Carlo
sampling) has an uncertainty of less than a factor of two (given as the
68\,\% coverage probability of the Monte-Carlo sampling) in the
astrophysically relevant temperature range.

Under typical astrophysical conditions of X-ray bursters (density $\rho
\approx 10^6$\,g/cm$^3$ and \al\ mass fraction $Y_\alpha \approx 0.27$) we find
that the rate of the $^{18}$Ne\rap $^{21}$Na reaction exceeds the
$\beta^+$-decay rate of $^{18}$Ne at $T_9 = 0.60 \pm 0.02$; i.e.,
the break-out temperature is very well constrained by the present work. At
this break-out temperature \Nsv\ is mainly determined by the 
resonance at 938\,keV with its newly confirmed $J^\pi = 1^-$ assignment
\cite{Zhang14}.

\acknowledgments
This work was supported by OTKA (K101328 and K108459) and by the
U.S. Department of Energy under Contract No. DE-FG02-97ER41041.

\appendix

\section{Fit of the reaction rate \Nsv }
\label{sec:rate_fit}
The recommended reaction rate \Nsv $_{\rm{rec}}$ is fitted by the usual
expression, see e.g.\ \cite{Rau00}, Eq.~(16):
\begin{eqnarray}
\frac{N_A < \sigma v >}{{\rm{cm}}^3 {\rm{s}}^{-1} {\rm{mol}}^{-1}} = 
  & & \, \, \exp{(a_0 + a_1 T_9^{-1} + a_2 T_9^{-1/3} + a_3 T_9^{1/3}} 
\nonumber \\ 
  & & \, \, \, \,
  + a_4 T_9 + a_5 T_9^{5/3} + a_6 \ln{T_9})
\label{eq:fit}
\end{eqnarray}
The $a_i$ parameters are listed in Table \ref{tab:fit}. The deviation of the
fitted rate is always below 12\,\% over the full temperature range $0.25
\le T_9 \le 3$ and typically far below 5\,\% in the most relevant range $1 \le
T_9 \le 2$; but the fit should not be used below $T_9 = 0.2$ or above $T_9 =
3.5$. 
\begin{table}[bth]
\caption{\label{tab:fit}
Fit parameters $a_i$ of the recommended reaction rate \Nsv
$_{\rm{rec}}$ from Eq.~(\ref{eq:fit}). The parametrization is valid
for the full temperature range $0.25 \le T_9 \le 3$ with a deviation of less
than 12\,\%. The fit formula in Eq.~(\ref{eq:fit}) should not be used below
$T_9 = 0.2$ or above $T_9 = 3.5$.
}
\begin{center}
\begin{tabular}{ccccccc}
\hline
\multicolumn{1}{c}{$a_0$}
& \multicolumn{1}{c}{$a_1$} 
& \multicolumn{1}{c}{$a_2$} 
& \multicolumn{1}{c}{$a_3$} 
& \multicolumn{1}{c}{$a_4$} 
& \multicolumn{1}{c}{$a_5$} 
& \multicolumn{1}{c}{$a_6$} \\
\hline
   $590.68$ 
& $-52.89$ 
& $1916.23$ 
& $-2586.60$
& $136.60$ 
& $-6.86$ 
& $1338.35$ \\
\hline
\end{tabular}
\end{center}
\end{table}

\end{document}